# Botnet Detection by Monitoring Similar Communication Patterns

Hossein Rouhani Zeidanloo
Faculty of Computer Science and Information System
University of Technology Malaysia
54100 Kuala Lumpur, Malaysia
h_rouhani@hotmail.com

Azizah Bt Abdul Manaf
College of Science and Technology
University of Technology Malaysia
54100 Kuala Lumpur, Malaysia
azizah07@citycampus.utm.my

*Abstract—* Botnet is most widespread and occurs commonly in today's cyber attacks, resulting in serious threats to our network assets and organization's properties. Botnets are collections of compromised computers (Bots) which are remotely controlled by its originator (BotMaster) under a common Command-and-Control (C&C) infrastructure. They are used to distribute commands to the Bots for malicious activities such as distributed denial-of-service (DDoS) attacks, spam and phishing. Most of the existing Botnet detection approaches concentrate only on particular Botnet command and control (C&C) protocols (e.g., IRC,HTTP) and structures (e.g., centralized), and can become ineffective as Botnets change their structure and C&C techniques. In this paper at first we provide taxonomy of Botnets C&C channels and evaluate well-known protocols which are being used in each of them. Then we proposed a new general detection framework which currently focuses on P2P based and IRC based Botnets. This proposed framework is based on definition of Botnets. Botnet has been defined as a group of bots that perform similar communication and malicious activity patterns within the same Botnet. The point that distinguishes our proposed detection framework from many other similar works is that there is no need for prior knowledge of Botnets such as Botnet signature.

## I. INTRODUCTION

Bot is a new type of malware [1] installed into a compromised computer which can be controlled remotely by BotMaster for executing some orders through the received commands. After the Bot code has been installed into the compromised computers, the computer becomes a Bot or Zombie [2]. Contrary to existing malware such as virus and worm which their main activities focus on attacking the infecting host, bots can receive commands from BotMaster and are used in distributed attack platform.

Botnets are networks consisting of large number of Bots. Botnets are created by the BotMaster(a person or a group of person which control remote Bots) to setup a private communication infrastructure which can be used for malicious activities such as Distributed Denial-of-Service(DDoS), sending large amount of SPAM or phishing mails, and other nefarious purpose [ 3,4,5 ]. Bots infect a person's computer in many ways. Bots usually disseminate themselves across the Internet by looking for vulnerable and unprotected computers to infect. When they find an unprotected computer, they infect it and then send a report to the BotMaster. The Bot stay hidden until they are announced by their BotMaster to perform an attack or task. Other ways in which attackers use to infect a computer in the Internet with Bot include sending email and using malicious websites, but common way is searching the Internet to look for vulnerable and unprotected computers [6] .

The main difference between Botnet and other kind of malwares is the existence of Command-and-Control (C&C) infrastructure. The C&C allows Bots to receive commands and malicious capabilities, as devoted by BotMaster. BotMaster must ensure that their C&C infrastructure is sufficiently robust to manage thousands of distributed Bots across the globe, as well as resisting any attempts to shutdown the Botnets. Recently, attackers are also continually improving their approaches to protect their Botnets. The first generation of Botnets utilized the IRC (Internet Relay Chat) channels as their Common-and-Control (C&C) centers. The centralized C&C mechanism of such Botnet has made them vulnerable to being detected and disabled. Therefore, new generation of Botnet which can hide their C&C communication have emerged, Peer-to-Peer (P2P) based Botnets. The P2P Botnets do not suffer from a single point of failure, because they do not have centralized C&C servers [12]. Attackers have accordingly developed a range of strategies and techniques to protect their C&C infrastructure. The rest of the paper is organized as follows. In Section 2, we analyze different Botnet communication topologies and consider the protocols that are currently being used in each model. In Section 3, we review the related work. In Section 4, we describe our proposed detection framework and all its components and finally conclude in section 5.

## II. BOTNET COMMUNICATION TOPOLOGIES

According to the Command-and-Control(C&C) channel, we categorized Botnet topologies into two different models, the Centralized model and the Decentralized model.

### A. Centralized model

In this model, one central point is in charge for exchanging commands and data between the BotMaster and Bots. In this model, BotMaster chooses a host (usually high bandwidth computer) to be the central point (Command-and-Control) server of all the Bots. The C&C server runs certain network services such as IRC or HTTP. The main advantage of this model is small message latency which cause BotMaster easily arranges Botnet and launch attacks. Since all connections happen through the C&C server, therefore, the C&C is a critical point in this model. In other words, C&C server is the weak point in this model. If somebody manages to discover and eliminates the C&C server, the entire Botnet will be useless and ineffective. Thus, it becomes the main negative aspect of this model.

Since IRC and HTTP are two common protocols that C&C server uses for communication, we consider Botnets in this model based on IRC and HTTP. Figure 1 shows the basic communication architecture for a Centralized model. There are two central points that forward commands and data between the BotMaster and his Bots.

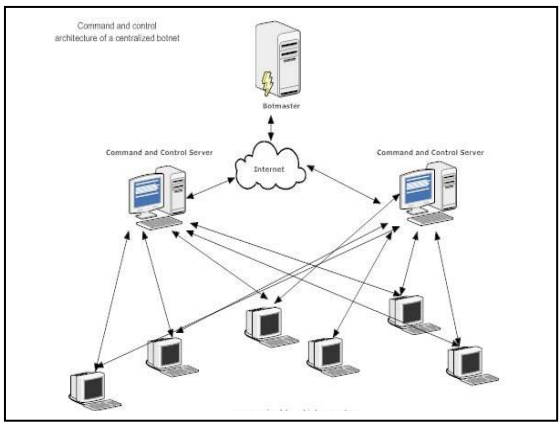

Figure 1. Command and control architecture of a Centralized model

*1) Botnet based on IRC :* The IRC is a form of real-time Internet text messaging or synchronous conferencing [13]. The protocol is based on the Client-Server model, which can be used on many computers in distributed networks. Some advantages which made IRC protocol widely being used in remote communication for Botnets are: (1) Low latency communication; (2) Anonymous real-time communication; (3) Ability of Group (many-to-many) and Private (one-to-one) communication; (4) simple to setup and (5) simple commands. The basic commands are connect to servers, join channels and post messages in the channels; (6) Very flexibility in communication. Therefore IRC protocol is still the most popular protocol being used in Botnet communication [5].

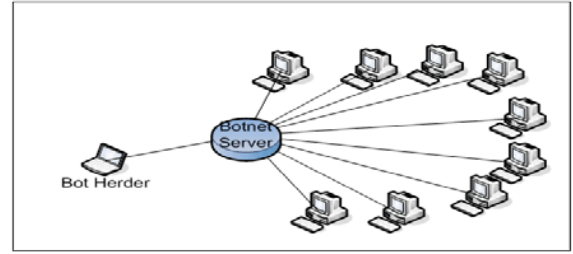

Figure 2. IRC based Botnet

In this model, BotMasters can command their Bots as a whole or command a few of the Bots selectively using one-to-one communication. The C&C server runs IRC service that is the same with other standard IRC service. BotMaster usually creates a designated channel on the C&C servers where all the Bots will connect, awaiting commands in the channel which will instruct each connected Bot to do the BotMaster's bidding. Figure 2 showed that there is one central IRC server that forwards commands and data between the BotHerder and his Bots.

*2) Botnet based on HTTP:* The HTTP protocol is another popular protocol used by Botnets. Since IRC protocol within Botnets became well-known, more internet security researchers gave attention to monitoring IRC traffic to detect Botnet. Consequently, attackers started to use HTTP protocol as a Command-and-Control communication channel to make Botnets become more difficult to detect. The main advantage of using the HTTP protocol is hiding Botnets traffics in normal web traffics, so it can easily bypasses firewalls with port-based filtering mechanisms and avoid IDS detection. Usually firewalls block incoming/outgoing traffic to unwanted ports, which often include the IRC port. There are some known Bots using the HTTP protocol, such as Bobax [16], ClickBot [13] and Rustock [17]. Gu et al in the reference [10] pointed out that the HTTP protocol is in a "pull" styleand the IRC is in a "push" style. However the architecture of both is same.

### B. Decentralized Model

Due to main disadvantage of Centralized model attackers started to build alternative Botnet communication system that is much harder to discover and to destroy. Hence, they decided to find a model in which the communication system does not completely depending on only some selected servers and even discovering and destroying a number of Bots. As a result, attackers exploit the idea of Peer-to-Peer (P2P) communication as a Command-and-Control(C&C) pattern which is more resilient to failure in the network. The P2P based C&C model will be used significantly in Botnets in the near future, and definitely Botnets that use P2P based C&C model impose much bigger challenge for defense of networks. Since P2P based communication is more robust than Centralized C&C communication, more Botnets will move to use P2P protocol for their communication.

In the P2P model, as shown in Figure 3, there is no Centralized point for communication. Each Bot keeps some connections to the other Bots of the Botnet. Bots act as both Clients and servers. A new Bot must know some addresses of the Botnet to connect there. If Bots in the Botnet are taken offline, the Botnet can still continue to operate under the control of BotMaster. P2P Botnets aim at removing or hiding the central point of failure which is the main weakness and vulnerability of Centralized model.

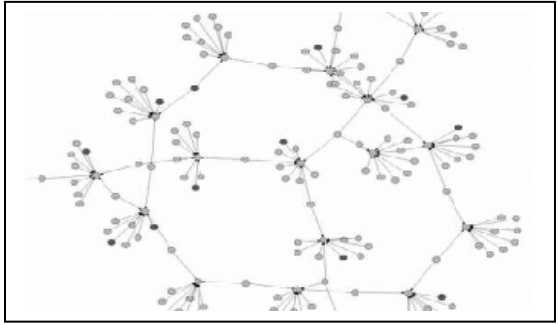

Figure 3. Example of Peer-to-peer Botnet Architecture

Some P2P Botnets operate to a certain extent decentralized and some completely decentralized. Those Botnets that are completely decentralized allow a BotMaster to inject a command into any Bots, and have it either be broadcasted to a specified node. Since P2P Botnets usually allow commands to be injected at any node in the network, the authentication of commands become essential to prevent other nodes from injecting incorrect commands.

## III. RELATED WORK

Different approaches have been proposed for detection of Botnet. There are essentially two approaches for botnt detection. One approach is based on locating honeynets in the network. And another approach is monitoring and analysis of passive network traffic [20].

There are many papers discussed how to apply honeynets for Botnet detection [5,3,21,22,23,24,1,25,26]. Honeynets are functional to understand Botnet characteristics and technology, but cannot detect bot infection all the times.

We can categorize passive network traffic monitoring approach to signature-based, anomaly-based, DNS-based and mining-based.

Signature-based Botnet detection technique uses the signatures of current Botnets for its detection. For instance, Snort [27] is capable to monitor network traffic to find signature of existing bots. Signature-based detection approach is only capable to be used for detection of well-known Botnets. Consequently, this solution is not functional for unknown bots.

Anomaly-based detection approaches try to detect Botnets based on a number of network traffic anomalies such as high network latency, high volumes of traffic, traffic on unusual ports, and unusual system behavior that could show existence of bots in the network [28]. Nevertheless this technique meets the problem of detecting unknown Botnets, but is not capable to realize an IRC network Botnet which has not been used yet for attacks. To solve this, Binkley and Singh [14] proposed an effective algorithm that combines TCP-based anomaly detection with IRC tokenization and IRC message statistics to create a system that can clearly detect client Botnets. This algorithm can also reveal bot servers [14]. However, Binkley's approach could be easily crushed by simply using a minor cipher to encode the IRC commands.

Lately, Gu et al. have proposed Botsniffer [13] that uses network-based anomaly detection to identify Botnet C&C channels in a local area network. Botsniffer is based on observation that bots within the same Botnet will likely reveal very strong similarities in their responses and activities. Therefore, it employs several correlation analysis algorithms to detect spatial-temporal correlation in network traffic with a very low false positive rate [13].

DNS-based detection techniques are based on DNS information generated by a Botnet. As mentioned before, bots normally begin connection with C&C server to get commands. In order to access the C&C server bots carry out DNS queries to locate the particular C&C server that is typically hosted by a DDNS (Dynamic DNS) provider. Therefore, it is feasible to detect Botnet DNS traffic by DNS monitoring and detect DNS traffic anomalies [29, 30].

In 2005, Dagon [31] proposed a method to discover Botnet C&C servers by detecting domain names with unusually high or temporally intense DDNS query rates. This method is similar to the approach proposed by Kristoff [32] in 2004.

In 2007, Choi et al. [29] suggested anomaly mechanism by monitoring group activities in DNS traffics. They defined some special features of DNS traffics to differentiate valid DNS queries from Botnet DNS queries. This method is more efficient than the prior approaches and can detect Botnet despite the type of bot by looking at their group activities in DNS traffic [29].

Geobl and Holz [15] proposed Rishi in 2007. Rishi is primarily based on passive traffic monitoring for odd or suspicious IRC nicknames, IRC servers, and uncommon server ports. They use n-gram analysis and a scoring system to detect bots that use uncommon communication channels, which are commonly not detected by classical intrusion detection systems [15]. The disadvantages of this method are that it cannot detect encrypted communication as well as non-IRC Botnets.

Strayer et al. [33] proposed a network-based approach for detecting Botnet traffic which used two step processes including separation of IRC flows at first, and then discover Botnet C&C traffic from normal IRC flows [33]. This technique is specific to IRC based Botnets.

Masud et al. [34] proposed effective flow-based Botnet traffic detection by mining multiple log files. They proposed several log correlation for C&C traffic detection. They categorize an entire flow to identify Botnet C&C traffic. This method can detect non-IRC Botnets[34].

Botminer [35] is the most recent approach which applies data mining techniques for Botnet C&C traffic detection. Botminer is an improvement of Botsniffer [13]. It clusters similar communication traffic and similar malicious traffic.

Then, it performs cross cluster correlation to identify the hosts that share both similar communication patterns and similar malicious activity patterns. Botminer is an advanced Botnet detection tool which is independent of Botnet protocol and structure. Botminer can detect real-world Botnets including IRC-based, HTTP-based, and P2P Botnets with a very low false positive rate [35].

As we mentioned above researches have proposed some approaches and techniques [14,15,16,13,17,18] for detecting Botnets. Majority of these approaches are developed for detecting IRC or HTTP based Botnets[14,15,18]. For instance, BotSniffer[13] is designed especially for detecting IRC and HTTP based Botnets. Rishi[15] is also desingned for detecting IRC based Botnets with using well-known IRC bot nickname patterns as signature. But recently we have witnessed that structure of Botnets moved from centralized to distributed (e.g., using P2P [9,19]). Consequently, the detection approaches designed for IRC or HTTP based Botnets may become ineffective against the new P2P based Botnets. Therefore, we need to develop a next generation Botnet detection system, which is also effective in the face of P2P based Botnets. In addition, we have to take into consideration that this detection system should require no prior knowledge of particular Botnets (such as Botnet signature, or C&C server names/addresses).

In order to come up with a new detection system that also meet the requirements for detection of P2P based Botnets, we studied the communication and activity characteristics of few P2P based Botnet( e.g. Storm Worm) and eventually come up with effective definition of Botnets; specially for P2P based Botnet:

"A group of bots (at least three) within the same Botnet will perform similar communication and malicious activities". Actually we share similar idea for definition of Botnet as proposed by Gu *et al.* in Botminer[35]. It means that if each bot within the same Botnet show different behavior (e.g. in terms of receiving instructions), the bots are only isolated infected systems that we cannot consider them as a Botnet based on our definition. According to definition above we proposed a new framework for detection of Botnets that mainly targets P2P based and IRC based Botnets, however the framework has the capability of adding another component for HTTP based Botnet detection. This framework monitors both the group of hosts that show similar communication pattern and performing malicious activities, and try to find common hosts on them.

### IV. PROPOSED BOTNET DETECTION FRAMEWORK AND COMPONENTS

Our proposed framework is based on passively monitoring network traffics. This model is based on the definition of P2P Botnets that multiple bots within the same Botnet will perform similar communication patterns and malicious activities. Figure 4 shows the architecture of our proposed Botnet detection system, which consist of 7 main components: Filtering, Application Classifier, Traffic Monitoring, Malicious Activity Detector, Analyzer, Monitoring & Clustering and Flows Analyzer.

Filtering is responsible to filter out unrelated traffic flows. The main benefit of this stage is reducing the traffic workload and makes application classifier process more efficient. Application classifier is responsible for separating IRC and HTTP traffics from the rest of traffics. Malicious activity detector is responsible to analyze the traffics carefully and try to detect malicious activities that internal host may perform and separate those hosts and send to next stage. Traffic Monitoring is responsible to detect the group of hosts that have similar behavior and communication pattern by inspecting network traffics. Analyzer is responsible for comparing the results of previous parts (Traffic Monitoring and Malicious Activity Detector) and finding hosts that are common on the results of both parts. Monitoring & Clustering is responsible to monitor the traffic flows and cluster the similar flows to same database. Flows Analyzer is responsible to detect the group of hosts that have similar behavior and communication patterns by comparing databases that received from previous stage for detecting IRC based bots.

*A. Filtering*

The main objective of Filtering is to reduce the traffic workload and makes the rest of the system perform more efficiently. Figure 5 shows the architecture of the filtering.

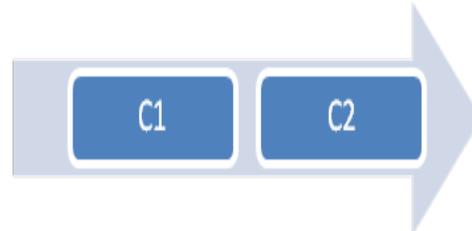

Figure 5. Traffics filtering stages

In C1, we filter out those traffics which targets (destination IP address) are recognized servers and will unlikely host Botnet C&C servers. For this purpose we used the top 500 websites on the web (Http://www.alexa.com/topsites), which the top 3 are google.com, facebook.com and yahoo.com.

In C2, we filter out handshaking processes (connection establishments) that are not completely established. Handshaking is an automated process of negotiation that dynamically sets parameters of a communications channel established between two entities before normal communication over the channel begins. It follows the physical establishment of the channel and precedes normal information transfer [36]. To establish a connection, TCP uses a three-way handshake; in this case we filter out the traffics that TCP handshaking have not completed. Like a host sends SYN packets without completing the TCP handshake. Based on our experience these flows are mostly caused by scanning activities.

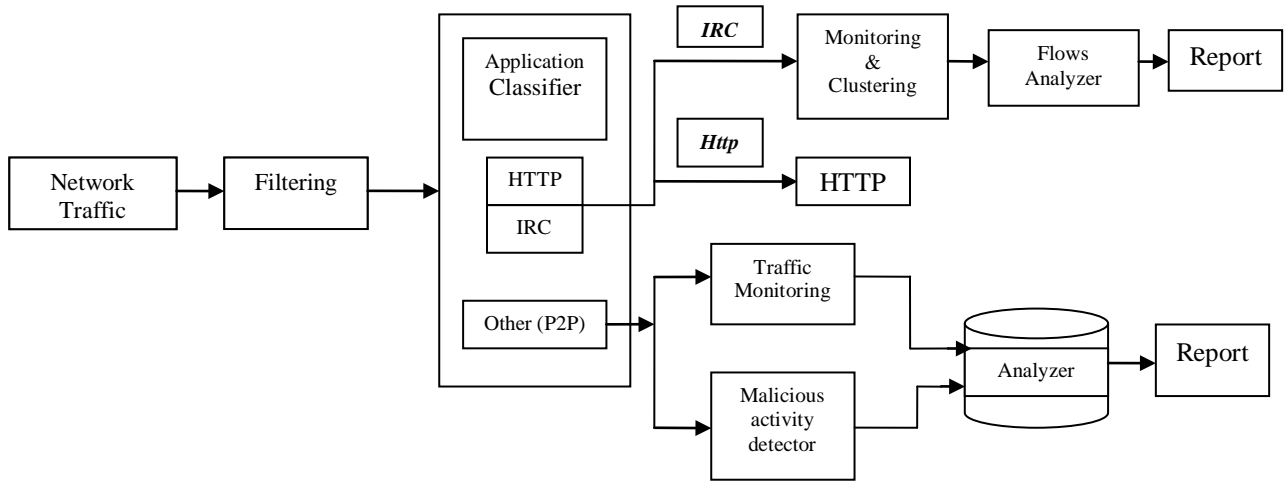

Figure 4. Architecture overview of our proposed detection framework

*B. Application classifier*

Application Classifier is responsible to separate IRC and HTTP traffics from the rest of traffics and send them to Monitoring & Clustering and HTTP component. For detecting IRC traffics we can inspect the contents of each packet and try to match the data against a set of user defined strings. For this purpose we use payload inspection that only inspects the first few bytes of the payload and looking for specific strings. These IRC specific strings are NICK for the client's nickname, PASS for a password, USER for the username, JOIN for joining a channel, OPER that says a regular user wants to become a channel operator and PRIVMSG that says the message is a private message [37]. Using this strategy for detecting IRC traffic is almost simple for most network intrusion detection software like Snort. In some cases botmasters are using encryption for securing their communication that make using packet content analysis strategy useless. This issue actually is not our target here.

In next step, we also have to separate Http traffics and send to Centralized part. For this purpose we also can inspect the first few bytes of Http request and if it has certain pattern or strings, separate it and send it to centralized part. For detecting Http traffics we focus on concept of Http protocol. Like most network protocols, HTTP uses the client-server model: An HTTP client opens a connection and sends a *request message* to an HTTP server (e.g. "Get me the file 'home.html'"); the server then returns a *response message*, usually containing the resource that was requested("Here's the file", followed by the file itself). After delivering the response, the server closes the connection (making HTTP a *stateless* protocol, i.e. not maintaining any connection information between transactions).[38]

In the format of Http request message, we are focusing on Http methods. Three common Http methods are "GET", "HEAD", or "POST": [38]

- A GET is the most common Http method; it says "give me this resource"

- A HEAD request is similar to GET request, except it asks the server to return the response headers only, and not the actual resource (i.e. no message body). This is helpful to consider characteristics of resources without downloading it which can help in saving bandwidth. We use HEAD when no need for a file's contents.
- A POST request is used to send data to the server to be processed in some way, like by a CGI script. A POST request is different from a GET request in the following ways:

  - There's a block of data sent with the request, in the message body. There are usually extra headers to describe this message body.
  
  - The *request URI* is not a resource to retrieve; it's usually a program to handle the data you're sending.
  
  - The HTTP response is normally program output, not a static file.

Therefore we inspect the traffics and if the first few bytes of an Http request contain "GET", "POST" or "HEAP", it's the indication of Http protocol and will separate those flows and send them to Centralized part. After filtering out Http and IRC traffics, the remaining traffics that have the probability of containing P2P traffics are send to Traffic Monitoring part and Malicious Activity Detector. However in parallel we can use other approaches for identifying P2P traffics. We have to take into consideration that P2P traffic is one of the most challenging application types. Identifying P2P traffic is difficult both because of the large number of proprietary p2p protocols, and also due to the deliberate use of random port numbers for communication. Payload-based classification approaches tailored to p2p traffic have been presented in [41, 40], while identification of p2p traffic through transport layer characteristics is proposed in [39]. Our suggestion for using specific application or tools for identifying P2P traffics other than sending remaining traffics is use of BLINC [42] that can identify general P2P traffics. In contrast to previous methods, BLINC is based on observing and identifying patterns of host behavior at the transport layer. BLINC investigates these patterns at three levels of increasing detail (i) the social, (ii) the functional and (iii) the application level. This approach has two important features. First, it operates in the dark, having (a) no access to packet payload, (b) no knowledge of port numbers and (c) no additional information other than what current flow collectors provide.[42]

### C. Traffic Monitoring

Traffic Monitoring is responsible to identify hosts that are likely part of Botnet during the time that hosts (bots) initiate attacks by analyzing flows characteristics and finding similarities among them. Therefore, we are capturing network flows and record some special information on each flow. We are using Audit Record Generation and Utilization System (ARGUS) which is an open source tool [43] for monitoring flows and record information that we need in this part. Each flow record has following information: Source IP(SIP) address, Destination IP(DIP) address, Source Port(SPORT), Destination Port(DPORT), Duration, Protocol(Pr), Number of packets($np$) and Number of bytes($nb$) transferred in both directions.

Table 1. Recorded information of network flows

| fi | SIP | DIP | Sport | Dport | Pr | $np$ | $nb$ | duration |
|---|---|---|---|---|---|---|---|---|
| f1 | | | | | | | | |
| f2 | | | | | | | | |
| . | | | | | | | | |
| . | | | | | | | | |
| fn | | | | | | | | |

Then we insert this information on a data base like Table 1, which $\{fi\}_{i=1…n}$ are network flows. After this stage we specify the period of time which is 6 hours and during each 6 hours, all n flows that have same Source IP, Destination IP, Destination port and same protocol (TCP or UDP) are marked and for each network flow {fi} (row) we calculate Average number of bytes per second and Average number of bytes per packet:

- Average number of bytes per second($nbps$) = Number of bytes/ Duration
- Average number of bytes per packet($nbpp$) = Number of Bytes/ Number of Packets

Then, we insert this two new values (nbps and $nbpp$) including SIP and DIP of the flows that have been marked into another database, similar to Table 2 . Therefore, during the specified period of time (6 hours), we might have a set of database, $\{fi\}_{i=1…m}$ which each of these databases have same SIP, DIP, DPORT and protocol (TCP/UDP). We are focusing just at TCP and UDP protocols in this part.

As we mentioned earlier, the bots belonging to the same Botnet have same characteristics. They have similar behavior and communication pattern, especially when they want to update their commands from botmasters or aim to attack a target; their similar behaviors are more obvious. Therefore, next step is to looking for groups of Databases that are similar to each other.

Table 2. Database for analogous flows

| SPort | DPort | *nbps* | *nbpp* |
|---|---|---|---|
| | | | |
| | | | |

We proposed a simple solution for finding similarities among group of databases. For each database we can draw a graph in x-y axis, which x-axis is the average number of bytes per packet (*nbpp*) and y-axis is average number of byte per second (*nbps*). (X, Y)= (bpp, bps).

For example, in database (di), for each row we have *nbpp* that specify x-coordinate and have *nbps* that determine y-coordinate. Both x-coordinate and y-coordinate determine a point (x,y) on the x-y axis graph. We do this procedure for all rows (network flows) of each database. At the end for each database we have number of points in the graph that by connecting those points to each other we have a curvy graph. We have an example, Figure 6, for two different databases based on data in our lab that their graphs are almost similar to each other.

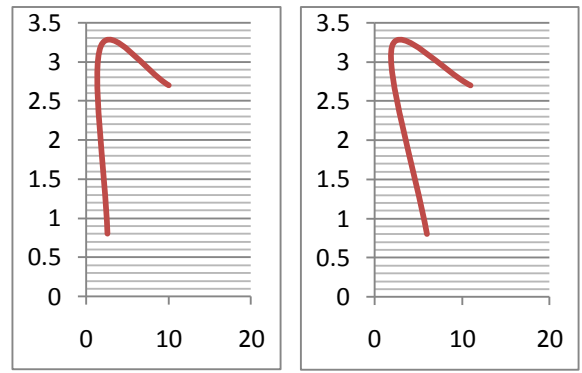

Figure 6: Example of two similar graphs based on data in our lab

Next step is comparing different x-y axis graphs, and during that period of time (each 6 hours) those graphs that are similar to each other are clustered in same category. The results will be some x-y axis graphs that are similar to each other. Each of these graphs is referring to their corresponding databases in previous step. We have to take record of SIP addresses of those hosts and send the list to next step for analyzing.

### D. Malicious Activity Detector

In this part we have to analyze the outbound traffic from the network and try to detect the possible malicious activities that the internal machines are performing. Each host may perform different kind of malicious activity but Scanning and Spamming are the most common and efficient malicious activities a botmaster may command their bots to perform [44,26,45]. The outputs of this part are the list of hosts which performed malicious activities.

*1) Scanning:* Scanning activities may be used for malware propagation and DOS attacks. Most scan detection has been based on detecting N events within a time interval of T seconds. This approach has the problem that once the window size is known, the attackers can easily evade detection by increasing their scanning interval. Snort are also use this approaches. Snort version 2.0.2 uses two preprocessors. The first is packet-oriented, focusing on detecting malformed packets used for "stealth scanning" by tools such as nmap [46]. The second is connection oriented. It checks whether a given source IP address touched more than X number of ports or Y number of IP addresses within Z seconds. Snort's parameters are tunable, but it suffers from the same drawbacks as Network Security Monitor(NSM)[47] since both rely on the same metrics [48]. Other work that are focusing on scan detection is by Staniford et al. on Stealthy Probing and Intrusion Correlation Engin( SPICE)[49]. SPICE is focusing on detecting stealthy scans, especially scans that spread across multiple source addresses and execute at very low rates. In SPICE there are anomaly scores for packets based on conditional probabilities derived from the SIP and DIP and ports. It uses simulated annealing to cluster packets together into port scan using heuristics that have developed from real scans[49]. An important need in our system is prompt response, however reaching to our goals which are promptness and accuracy in detecting malicious scanners is a difficult task. Another solution is also using Threshold Random Walk(TRW)[48], an online detection algorithm. TRW is based on sequential hypothesis testing.

After assessing different approaches for detecting scanning activities, the best solution for using in this part is Statistical sCan Anomaly Detection Engine( SCADE)[16], a snort processor plug-in system which has two modules, one for inbound scan detection and another one for detecting outbound attack propagation.

*a) Inbound Scan Detection(ISD):* In this part SCADE has focused on detection of scan activities based on ports that are usually used by malware. One of the good advantages of this procedure is that it is less vulnerable to DOS attacks, mainly because its memory trackers do not maintain per-external-source-IP. SCADE here just tracks scans that are targeted to internal hosts. The bases of Inbound Scan Detection are on failed connection attempts. SCADE in this part has defined two types of ports: High-Severity (hs) ports which representing highly vulnerable and commonly exploited services and low-severity (ls) ports. For make it more applicable in current situation SCADE focused on TCP and UDP ports as high-secure and all other as low-secure ports. There are different weights to a failed scan attempt for different types of ports.

The warning for ISD for a local host is produced based on an anomaly score that is calculated as based on this formula:

$$S = (w_1 fhs + w_2 fls)$$

fhs: indicate numbers of failed attempts at high-severity ports
fls: shows numbers of failed attempts at low-severity ports

*b) Outbound Scan Detection (OSD):* OSD is based on a voting scheme (AND, OR or MAJORITY). SCADE in this part has three parallel anomaly detection models that track all outbound connection per internal host:
• Outbound scan rate ($s_1$): Detects local hosts that perform high-rate scans for many external addresses.
• Outbound connection failure rate ($s_2$): Detects unusually high connection fail rates, with sensitivity to HS port usage. The anomaly score $s_2$ is calculated based on this formula:

$$S_2 = (w_1 fhs + w_2 fls)/C$$

fhs: indicate numbers of failed attempts at high-severity ports
fls: shows numbers of failed attempts at low-severity ports
C: is the total number of scans from the host within a time window.

• Normalized entropy of scan target distribution ($s_3$): Calculates a Zipf (power-law) distribution of outbound address connection patterns. A consistently distributed scan target model provides an indication of a possible outbound scan. It is used an anomaly scoring technique based on normalized entropy to identify such candidates:

$$S_3 = H/\ln(m)$$

H: is the entropy of scan target distribution

$$H = -\sum p_i \ln(p_i)$$

*m*: is the total number of scan targets
*pi*: is the percentage of the scans at target i

*2) Spam-related Activities:* E-mail spam, known as Unsolicited Bulk Email (UBE), junk mail, is the practice of sending unwanted email messages, in large quantities to an indiscriminate set of recipients. More than 95% of email on the internet is spam[50], which most of these spam are sent from Botnets. A number of famous Botnets which have been used specially for sending spam are Storm Worm which is P2P Botnet and Bobax that used Http as its C&C.
A common approach for detecting spam is the use of DNS Black/Black Hole List (DNSBL) such as (http://www.dnsbl.info/dnsbl-list.php). DNSBLs specify a list of spam senders' IP addresses and SMTP servers are blocking the mail according to this list. This method is not efficient for bot-infected hosts, because legitimate IP addresses may be used for sending spam in our network. Creation or misuse of SMTP mail relays for spam is one of the most well-known exploitation of Botnets. As we know user-level client mail application use SMTP for sending messages to mail server for relaying. However for receiving messages, client application usually use Post Office Protocol (POP) or the Internet Message Access Protocol (IMAP) to access the mail box on a mail server. Our idea in this part is very simple and efficient.

Our target here is not recognizing which email message is spam, though for detecting group of bots that sending spam with detecting similarities among their actions and behaviors. Therefore the content of emails from internal network to external network is not important in our solution. All we want to do is determining which clients have been infected by bot and are sending spam. For reaching to this target, we are focusing on the number of emails sending by clients to different mail servers. Based on our experience in our lab, using different external mail servers for many times by same client is an indication of possible malicious activities. It means that it is unusual that a client in our network send many emails to the same mail server (SMTP server) in the period of time like one day. Therefore, we are inspecting outgoing traffic from our network( gateway), and recording SIP and DIP of those traffics that destination ports are 25( SMTP) or 587(Submission) in the database. Based on network flows between internal hosts and external computers( SIP belong to mail servers) and the number of times that it can happen we can conclude which internal host is behaving unusual and are sending many emails to different or same mail servers.

*E. Analyzer*

Analyzer which is the last part of our proposed framework for detection of Botnets, is responsible for finding common hosts that appeared in the results of previous parts (Traffic Monitoring and Malicious Activity Detector).

*F. Monitoring and Clustering*

Since the architecture of communication between IRC server and bots is one-to-many (multicast) model, thus; the network flows to all bots should show similar characteristic and pattern. Our objective in this part is detection of IRC based Botnet by monitoring network traffics. Our approach in this part is based on identifying hosts that are likely part of a Botnet before initiating an attack, particularly during the time that IRC server commanding or updating their bots.

Monitoring & Clustering is responsible to inspect network traffics and clustering the similar characteristics of network flows. Consequently, we are capturing network flows and record some special information in each flow. We are using ARGUS for monitoring flows and record information that we need in this part. Each flow record has following information: Source IP(SIP) address, Destination IP(DIP) address, Source Port(SPORT), Destination Port(DPORT), Duration(Dr), Protocol(Pr), Packet Arrival Time(PAT), Number of packets (*np*) and Number of bytes (*nb*) transferred in both directions. Then, we insert this information in a data base as shown in Table 3, in which {fi}i=1…n are network flows.

After this stage, we specify the period of time which is 6 hours and during each 6 hours, all n flows that have same Source IP, Destination IP, Source port, Destination port, Packet Arrival Time (PAT) and same protocol (TCP or UDP), are marked and then for each network flow (row) we calculate *nbps* and *nbpp* based on formula that mentioned earlier.

Table 3. Recorded information of network flows

| fi | SIP | DIP | Sport | Dport | Dr | Pr | PAT | *np* | *nb* |
|---|---|---|---|---|---|---|---|---|---|
| f1 | | | | | | | | | |
| f2 | | | | | | | | | |
| . | | | | | | | | | |
| . | | | | | | | | | |
| fn | | | | | | | | | |

Then, we insert these two new values (*nbps* and *nbpp*) including SIP and DIP of the flows that have been marked into another database, similar to Table 2. Therefore, during the specified period of time (6 hours), we might have a set of database,{di}i=1…m in which each of these databases has the same SIP, DIP, DPORT, PAT and protocol (TCP/UDP). We are focusing just at TCP and UDP protocols in this part. These databases are sent to next stage, Flows analyzer, for finding similar databases.

*G. Flows Analyzer*

Flows Analyzer is responsible for looking a group of databases that are similar to each other. The comparison and finding analogous databases is similar to approach that has been described in Traffic Monitoring component. After finding similar databases we have to take a record of SIP addresses of those hosts and send them as a group of bot that are belong to IRC based Botnet.

V. CONCLUSION

The first seminar on Botnets was hold in 2007 and since then many Botnet detection techniques have been proposed and also some real bot detection systems have been implemented (e.g. BotHunter by Gu et al. [16]). Botnet detection is a challenging problem. In this paper at first we have defined taxonomy for better understanding of Botnets. Then we proposed a new general detection framework which currently focuses on P2P based Botnets and IRC based Botnets. This proposed framework is based on the definition of Botnets. Botnets have been defined as a group of bots that will perform similar communication and malicious activities pattern within the same Botnet. The point that distinguishes our proposed detection framework from many other similar works is that there is no need for prior knowledge of Botnets such as Botnet signature. In addition, we plan to further improve the efficiency of our proposed detection framework with adding unique detection method in HTTP part and make it as one general system for detection of Botnet and try to implement it in near future.

ACKNOWLEDGMENT

The authors would like to express their appreciation to Universiti Teknologi Malaysia (UTM) for their invaluable supports (technically & financially) in encouraging the authors to publish this paper.

AUTHORS PROFILE

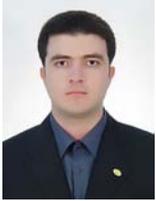

Hossein Rouhani Zeidanloo received his B.Sc. in software engineering from Meybod University, Iran. He is currently completing his Master's degree in Information Security at the Universiti Teknologi Malaysia (UTM). He has published two papers in international journals and also published many papers in international conferences around the world. His area of interest is Network Security and Ethical Hacking.

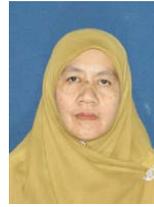

Azizah Abdul Manaf is a Professor at Universiti Teknology Malaysia (UTM). She graduated with B. Eng. (Electrical) 1980, MSc. Computer Science (1985) and PhD in 1995 from UTM. Her current areas of interest and research are image processing, watermarking, steganography, Information Security, Botnets and Worm, Intrusion Detection and computer forensics and have postgraduate students at the Masters and PhD level to assist her in these research areas. She has written numerous articles in journals and presented an extensive amount of papers at national and international conferences on her research areas. Prof. Dr. Azizah has also held management positions at the University and Faculty level such as Head of Department, Deputy Dean, Deputy Director and Academic Director pertaining to academic development as well as on training for teaching and learning methodologies at UTM.